\documentclass[10pt,twocolumn,showpacs,superscriptaddress,preprintnumbers,floatfix]{revtex4}
\pdfoutput=1
\usepackage{latexsym}
\usepackage{amsmath}
\usepackage{amsfonts}
\usepackage{graphicx}

\def\laq{~\raise 0.4ex\hbox{$<$}\kern -0.8em\lower 0.62
ex\hbox{$\sim$}~}
\def\gaq{~\raise 0.4ex\hbox{$>$}\kern -0.7em\lower 0.62
ex\hbox{$\sim$}~}

\def\beq{\begin{equation}}
\def\eeq{\end{equation}}
\def\bea{\begin{eqnarray}}
\def\eea{\end{eqnarray}}
\def\bean{\begin{eqnarray*}}
\def\eean{\end{eqnarray*}}

\def\re {\rangle}
\def\eff{e\!f\!f}

\def \a {\alpha}

   \def\be{\begin{equation}}
   \def\ee{\end{equation}}
   \def\ba{\begin{eqnarray}}
   \def\ea{\end{eqnarray}}

\begin{document}
\addtolength{\belowdisplayskip}{-0.0cm}
\addtolength{\abovedisplayskip}{-0.0cm}

\title{Gauge invariant backreaction in general single field models of inflation}

\author{G. Marozzi}
\affiliation{Universit\'e de Gen\`eve, D\'epartement de Physique Th\'eorique and CAP \\
       24 quai Ernest-Ansermet, CH-1211 Gen\`eve 4, Switzerland}
\author{G. P. Vacca}
\affiliation{INFN, sezione di Bologna, via Irnerio 46, I-40126 Bologna, Italy}


\begin{abstract} 

In a general single field inflationary model we consider the effects of long wavelength scalar fluctuations on the effective expansion rate 
and equation of state seen by a class of free falling observers, using  a physical gauge invariant formulation.
In a previous work we showed that for a free massive inflaton no backreaction is observed within some constraints. 
In this paper we extend the validity of our previous results to the case of an arbitrary self-interacting inflation potential,
working to second order in cosmological perturbation theory and to all order in slow-roll approximation. 
For these general inflationary models, we also show the equivalence of the free falling observers to the ones comoving with the inflaton field.
\end{abstract}

\maketitle

\section{Introduction}
The problem of computing the quantum backreaction induced by cosmological fluctuations
in an inflationary era has a long history~\cite{ABM} and has been rather controversial~\cite{Uc,All}.
With the purpose of settling such controversial issue, 
we have recently performed an  investigation~\cite{FMVVback} based on  a new covariant and gauge invariant (GI) approach~\cite{GMV1,GMV2}. 
In \cite{FMVVback}  we analysed, for a model with a massive minimally coupled free single inflaton field, how several observers see the effective
expansion rate and the effective equation of state of the Universe.
A first (trivial) observation in \cite{FMVVback}, independent from the particular model of inflation investigated in that letter, 
was that an observer which sees as unperturbed his/her spatial hypersurface (i.e. the class of observers associated to the uniform curvature gauge)
experiences exactly zero backreaction effects, to all order in perturbation theory and without the needed of the long wavelength
approximation, having an effective scale factor equal to its homogeneous value~\footnote{As also underlined in \cite{BLC}, this result corrects a previous statement \cite{BBM} that no gauge with identically zero backreaction can be found.}.
The main achievement obtained in \cite{FMVVback} was that, in the model with a massive minimally coupled free single inflaton field considered,
the free falling observers do not experience any quantum backreaction.
In particular, the analysis took into account the scalar quantum cosmological fluctuations,  in the long wavelength (LW) approximation and
to first order in the slow-roll approximation,  
derived within a fully second order perturbation theory expansion.
On the other hand we showed later that in such a model other special observers, such as the isotropic ones~\cite{MV1} and
the ones associated to a second light test field~\cite{MVB}, may observe quantum effects on the effective expansion rate and equation of state.
Effects that could shorten the observed phase of inflation \cite{MV1,MVB} and change deeply the "apparent" inflationary dynamics \cite{MVB}.
In this paper we want to extend the investigation of \cite{FMVVback} to a more general cosmological model, with one minimally coupled, generically self interacting, inflaton field, and to all order in slow-roll approximation.
We shall analyse in particular the free falling (geodetic) observers.

Let us review very briefly the basic elements of the general approach.
One can construct a general nonlocal observable performing
quantum averages of a scalar field $S(x)$ in a spacetime region which is a hypersurface
$\Sigma_{A_0}=\{x | A(x)=A_0\}$ defined by a scalar field $A(x)$ with a timelike gradient.
In particular we can adopt a GI definition, which in the (barred) coordinate system 
$\bar{x}^{\mu}=(\bar{t}, \vec{x})$, where the scalar $A$ is homogeneous, corresponds to the following quantity \cite{GMV1}
\beq
\langle S \rangle_{A_0}={\langle \sqrt{|\overline{\gamma}(t_0, {\vec{x}})|} 
\,~ \overline{S}(t_0, {\vec{x}}) \re \over  \langle   
\sqrt{|\overline{\gamma}(t_0, {\vec{x}})|} \re} \,,
\label{media}
\eeq
where $\overline{\gamma}(t_0, {\vec{x}})$  is the
determinant of the induced three dimensional metric on $\Sigma_{A_0}$.
The natural foliation of spacetime is then defined by the four vector 
\be
n^\mu= -\frac{\partial^\mu A}{(-\partial^\nu A\partial_\nu A)^{1/2}}\,,
\ee
which characterizes completely the physical properties of the class of observers sitting on the hypersurface $A(x)=A_0$ \cite{Mar}. 

Here we are interested in the dynamics encoded in the effective scale factor
$a_{eff}=\langle\sqrt{|\bar{\gamma}|}\, \rangle ^{1/3}$. 
In particular, the GI definition of the associated expansion rate is given by~\cite{GMV2} 
\be
\left(\frac{1}{a_{eff}}\frac{\partial \, a_{eff}}{\partial A_0} \right)^2
=\, \,\frac{1}{9} \left\langle\frac{\Theta}{(-\partial^\mu A\partial_\mu A)^{1/2}} 
\right\rangle_{A_0}^2\,,
\label{genEQ}
\ee
where  $\Theta=\nabla_\mu n^\mu$ is the expansion scalar of the timelike congruence $n^\mu$.
We can then extract the dynamical informations solving the Einstein and matter equations of motion in any gauge.

We shall work in the context of a spatially flat FLRW background
geometry. In order to deal with the metric components in any specific frame
we expand our background fields $\{g_{\mu\nu}\}$
up to second order in the non-homogeneous perturbations as follows:
\bea
g_{00}\!\!&=&\!\! -1\!-\!2 \a\!-\!2 \a^{(2)}\,, \,\,  
g_{i0}=-{a\over2}\!\left(\beta_{,i}\!+\!B_i \right) \!
-\!{a\over2}\!\left(\!\beta^{(2)}_{,i}\!+\!B^{(2)}_i\!\right) 
\nonumber\\
\!\!\!\!g_{ij}\!\! &=&\!\!  a^2 \!\Bigl[ \delta_{ij} \! 
\left( \!1\!-\!2 \psi\!-\!2 \psi^{(2)}\right)
+D_{ij} (E+E^{(2)})
\nonumber\\
& & \!\!\!\!\!\!\!\!\!\!
+{1\over 2} \left(\chi_{i,j}+\chi_{j,i}+h_{ij}\right)
+ {1\over 2} \left(\chi^{(2)}_{i,j}+\chi^{(2)}_{j,i}+h^{(2)}_{ij}\right)\Bigr]
\label{GeneralGauge}
\eea
where $D_{ij}=\partial_i \partial_j- \delta_{ij} (\nabla^2/3)$ and, for notational
simplicity, we remove an upper script in first order quantities. 
$\alpha$, $\beta$, $\psi$, $E$ are pure scalar perturbations, 
$B_i$ and $\chi_i$ are transverse vectors ($\partial^i B_i=0$ and 
$\partial^i \chi_i=0$), $h_{ij}$ is a traceless and transverse tensor 
($ \partial^i h_{ij}=0=h^{i}_i$), and the same notation applies to
the case of the second-order perturbations.
 
In a single field model of inflation,
the Einstein equations connect those fluctuations directly with the inflaton ones. 
In particular the inflaton field can be written to second order as
$\Phi(x)=\phi(t)+\varphi(x)+\varphi^{(2)}(x)$.

Since in Eq.(\ref{GeneralGauge}) there are $10$ degrees of freedom, which are in part redundant, a
gauge fixing is required.
In general the metric can be gauge fixed eliminating two scalar and one vector perturbations. 
The most popular gauge choices are 
the synchronous gauge (SG), defined by $g_{00}=-1$ and $g_{i0}=0$,
the uniform field gauge (UFG), defined by setting $\Phi(x)=\phi(t)$ and 
by other conditions (we consider $g_{i0}=0$), and 
the uniform curvature gauge (UCG), defined by
$g_{ij}=a^2\left[\delta_{ij}+\frac{1}{2} \left(h_{ij}+h^{(2)}_{ij}\right)\right]$.

The inflationary model here considered is a model with a self interacting
minimally coupled inflaton scalar field $\Phi$ with potential $V$:
   \be
    S = \int d^4x \sqrt{-g} \left[ 
\frac{R}{16{\pi}G}
    - \frac{1}{2} g^{\mu \nu}
    \partial_{\mu} \Phi \partial_{\nu} \Phi - V(\Phi) \right] \,.
    \label{action}
    \ee

\section{Backreaction for free falling observers}
To second order in perturbation theory and in the LW approximation, 
we have then the following result for the backreaction on the expansion rate as seen from an observer 
sitting on a particular hypersurface (the one correspondent to the barred gauge defined in the previous section)
\cite{FMVVback}
\bea
H_{\eff}^2 &\equiv& \dot{A}^{(0) 2} \left(\frac{1}{a_{\eff}}\frac{\partial \, a_{\eff}}{\partial A_0} \right)^2 \nonumber\\
&=&H^2 \!\left[1+\frac{2}{H}\langle \bar{\psi}\dot{\bar{\psi}}\rangle-
\frac{2}{H}\langle \dot{\bar {\psi}}^{(2)}\rangle\right]\,.
\label{Heff}
\eea
Let us note that we have neglected, in our computations, the dependence on the tensor perturbations.

Let us now consider a class of free falling observers and perform the calculation in the UCG.
In such a gauge and in the LW limit we have 
(see \cite{FMVVback}):
\begin{eqnarray}
\bar{\psi}\!\! &=&\!\! H\!\! \int^t \!\!dt' \alpha\,, 
\nonumber \\
\bar{\psi}^{(2)} \!\!&=&\!\! -H \alpha\! \int^t \!\!dt' \alpha-\frac{\dot{H}+2 H^2}{2} 
\left[\int^t \!\!dt' \alpha\right]^2\nonumber\\
&& +\frac{H}{2} \int^t \!\!dt' \left(2 \alpha^{(2)}-\alpha^2\right)\,.
\end{eqnarray} 

In order to evaluate the backreaction via Eq.(\ref{Heff}), to all order in slow-roll approximation but in the LW limit, 
we need the relations between the inflaton and the metric fluctuations
in the UCG.
To first order, one has
\be
\alpha=\frac{1}{2M^2_{Pl}}\frac{\dot{\phi}}{H} \varphi \,,
\label{Eq_usefull1}
\ee
where $M_{Pl}^{-2}=8\pi G$.
Solving the first order equation of motion for the inflaton field
\be 
\ddot{\varphi}+3 H \dot{\varphi}+\left[V_{\phi\phi}+2 \frac{d}{dt} \left(3 H+\frac{\dot{H}}{H}\right)\right]\varphi=0
\ee
in the LW approximation, but to all orders in the slow-roll approximation, one gets the well known solution
\be
\varphi=f(\vec{x}) \frac{\dot{\phi}}{H}  \Rightarrow
\int^t \!dt' \alpha=-f(\vec{x}) \int^t \! dt'
\frac{\dot{H}}{H^2}=\frac{1}{\dot{\phi}} \varphi\,. 
\label{Eq_usefull2}
\ee
This leads to
\be 
\bar{\psi}=\frac{H}{\dot{\phi}} \varphi=f(\vec{x}) \Rightarrow \dot{\bar{\psi}}=0\,.
\ee
On the other hand, to second order we obtain
\be 
\bar{\psi}^{(2)} = -  \frac{1}{2 M^2_{Pl}} \left(\frac{1}{2}-\frac{H^2}{\dot{H}}\right) \varphi^2 +
\frac{H}{2} \int^t dt' \left(2 \alpha^{(2)}-\alpha^2\right)\,.
\label{psi2}
\ee
The question is whether or not also $\dot{\bar{\psi}}^{(2)} =0$ in the LW limit, independently from the potential $V(\phi)$ and to all order 
in the slow-roll approximation.

Let us consider the equation of motion of $\alpha^{(2)}$ in the UCG.
After some algebra, from the general expressions given in Eqs. (24) and (25) of \cite{FMVV_II}, one can obtain, in the LW limit, the simplified form
\be 
\alpha^{(2)}=\frac{1}{2 M_{Pl}^2} \frac{\dot{\phi}}{H} \varphi^{(2)} +\frac{1}{2 M_{Pl}^2} \left( -\frac{3}{2} \frac{\dot{H}}{H^2}
+\frac{1}{2 H} \frac{\ddot{\phi}}{\dot{\phi}} \right) \varphi^2\,,
\label{alpha2}
\ee
which still depends on $\varphi^{(2)}$. 
The dynamics of $\varphi^{(2)}$ can be obtained in the UCG and in the LW limit, 
using Eqs. (26) and (27) of ~\cite{FMVV_II}, from the condition $\nabla^2 \beta^{(2)}\equiv 0$. 
After several algebraic manipulations,  and using the relation $V=M_{Pl}^2 (3 H^2 + \dot{H})$, one finds
\be
\frac{H M_{Pl}^2}{\dot{\phi}} \varphi^{(2)}= \!\int^t \!\!\!dt' \!\left[ -\frac{\dot{H}}{2H}+\frac{H}{8 \dot{H}}
\frac{\ddot{H}^2}{\dot{H}^2}\!\!+\frac{3}{8}\frac{\ddot{H}}{\dot{H}}-\frac{H{H}^{(3)}}{8\dot{H}^2}\right]\!\varphi^2\,,
 \label{Lastphi2}
\ee
where ${H}^{(3)}=\frac{d^3}{dt^3} H$.

Using this last relation \eqref{Lastphi2} in Eq.(\ref{alpha2}) we can obtain the following result, valid in the LW limit for a generic scalar field potential,
\be 
\alpha^{(2)}=\left(-\frac{\dot{H}}{H^2}+\frac{1}{4} \frac{\ddot{H}}{\dot{H} H}\right)\frac{\varphi^2}{M_{Pl}^2}.
\ee
Indeed, taking into account the time dependence of $\varphi$ in the LW limit, one can perform exactly the integral in Eq.(\ref{Lastphi2}),
thanks to the following useful identities
\be 
\int^t \!\!dt' \frac{\dot{H}^2}{H^3}=-\frac{\dot{H}}{2 H}+\int^t dt' \frac{\ddot{H}}{2 H^2}\nonumber\,,
\ee
\be 
\int^t \!\!dt' \left[ \frac{\ddot{H}}{H^2}+\frac{1}{H}\left(\frac{\ddot{H}}{\dot{H}}\right)^2\!\!-\frac{H^{(3)}}{H \dot{H}}\right]=
-\frac{\ddot{H}}{H \dot{H}}\nonumber\,.
\ee
Going back to Eq.(\ref{psi2}) we obtain
\bea
\bar{\psi}^{(2)} &=& \left(-\frac{1}{4}+\frac{1}{2} \frac{H^2}{\dot{H}}\right) \frac{\varphi^2}{M_{Pl}^2} \nonumber\\
&{}&+\frac{H}{2 M_{Pl}^2} \int^t \!\!dt' \left(- \frac{3}{2} \frac{\dot{H}}{H^2}+\frac{1}{2} \frac{\ddot{H}}{H \dot{H}}\right)\varphi^2
\label{psi2B}
\eea
and, again, solving the integral in the LW limit one finally gets
\be 
\bar{\psi}^{(2)} =\frac{H^2}{2 \dot{H}} \frac{\varphi^2}{M_{Pl}^2}=-\bar{\psi}^2=- f(\vec{x})^2\,,
\ee
namely $\dot{\bar{\psi}}^{(2)}\equiv 0$ in the LW limit.

We are now able to evaluate trivially the effective expansion rate in Eq.~\eqref{Heff}, since we have shown that both first and second order fluctuation
$\bar{\psi}$ and $\bar{\psi}^{(2)}$ are constant in time so that we have
 \be
H_{\eff}^{\rm Free \,Falling}\equiv \dot{A}^{(0)\,2}  \left(\frac{1}{a_{\eff}}\frac{\partial \, a_{\eff}}{\partial A_0} \right)= H\,.
\label{HeffFinal}
\ee
Therefore the geodetic observers do not see on the effective Hubble rate any quantum backreaction effect,
in the LW limit, for any potential $V(\phi)$ and to all orders in slow-roll parameters.
This fact extends the domain of validity of our previous results~\cite{FMVVback}, since previously the derivation was obtained
only for a quadratic potential corresponding to a free massive scalar field and not analysed to all orders in the slow-roll parameters.

Let us conclude this section by noting that one can define the effective observers dependent energy density $\rho_{\eff}$
by 
\be 
\dot{A}^{(0) 2}  \left(\frac{1}{a_{eff}}\frac{\partial \, a_{eff}}{\partial A_0} \right)^2=\frac{8 \pi G}{3} \rho_{\eff} \,,
\ee
while the effective pressure $p_{\eff}$ 
can be obtained by 
\be 
-\frac{1}{a_{\eff}}\dot{A}^{(0)}  \frac{\partial}{\partial A_0}\left( \dot{A}^{(0)}  \frac{\partial}{\partial A_0} a_{\eff}\right)=\frac{4 \pi G}{3} (\rho_{\eff}+3\, p_{eff}) \,.
\ee 
Trivially the effective equation of state will be given by $w_{eff}=p_{eff}/\rho_{eff}$ and, 
following \cite{MV1}, we obtain an exactly zero backreaction on such effective 
equation of state as seen from the class of observers considered (the free falling ones). 
The dynamic of any single field model is unchanged when a free falling observer is 
taken as reference point.

\section{Comoving observers}

Let us now further show that the general properties proved above are also true for a class of observers comoving with the inflaton field $\Phi(\vec{x}, t)$. Namely, that in a general one-field inflationary model
the free falling observers and the comoving observers are equivalent.
This point was already shown in \cite{FMVVback}.
Here we want to give a more general and explicit proof, valid in a generic single field model, showing 
that the two scalars associated to the two observers
(the one defining the free-falling observers and the one defining the observers comoving with the inflaton field) 
do coincide in the LW limit for an arbitrary potential $V(\Phi)$ and to all order in slow-roll approximation up to second order in cosmological perturbation theory.

Following \cite{Mar} the general scalar $C$, which is used to define an observer comoving with the inflaton field (and therefore by definition homogeneous in the UFG), is given in an arbitrary coordinate system by
\be
C(x)=C^{(0)}+\frac{\dot{C}^{(0)}}{\dot{\phi}} \varphi 
+\frac{\dot{C}^{(0)}}{\dot{\phi}} \varphi^{(2)}
+\frac{\dot{C}^{(0)}}{2 \dot{\phi}^{2}} \left(
\frac{\ddot{C}^{(0)}}{\dot{C}^{(0)}}-
\frac{\ddot{\phi}}{\dot{\phi}} \right)\varphi^{2} \,.
\label{UFGsc}
\ee
On the other hand the geodetic, or free falling, observer is associated to a scalar field homogeneous in the SG.
This can be written in an arbitrary reference system as $A(x)=A^{(0)}(t)+A^{(1)}(x)+A^{(2)}(x)$, with \cite{Mar}
\begin{eqnarray}
\!\!\!\!A^{(1)}(x)&=& \!\!\dot{A}^{(0)} \int dt \,\alpha^{(1)} \nonumber \\
\!\!\!\!A^{(2)}(x)&=& \!\!\frac{1}{2} \ddot{A}^{(0)} \left(\int dt\, \alpha^{(1)}\right)^2\nonumber\\
& &\!\!+\dot{A}^{(0)} \!\!\int \!\!dt \Biggl[-\frac{1}{2} \alpha^{(1)\,2}+ 
\frac{1}{2 a^2} \left(\int \!\!dt \vec{\nabla} \alpha^{(1)}\right)^2\!\!\nonumber\\
& &\!\!-\frac{1}{2 a}
\left(\int dt \, \alpha^{(1)}_{,j}\right)
\left(\beta^{(1),j}+B^{(1)\,j}\right)+\alpha^{(2)}\nonumber\\
& &\!\!+\frac{1}{8}
\left(\beta^{(1)}_{,j}+B^{(1)}_j\right)
\left(\beta^{(1),j}+B^{(1)\,j}\right)  
\Biggr]\,.
\label{A_SG}
\end{eqnarray}
As seen, the physical properties of these classes of observers,
associated to different hypersurfaces, 
are encoded in the vector fields normal to the correspondent hypersurfaces.
As shown in \cite{Mar} these vectors are independent from the background homogeneous value of the 
correspondent scalars. For the sake of simplicity we therefore take $A^{(0)}(t)=C^{(0)}(t)=\phi(t)$. The first consequence of this choice is
the simplification of Eq.(\ref{UFGsc}), which becomes
\be 
C(x)=\Phi(x)=\phi(t)+\varphi(x)+\varphi^{(2)}(x)\,.
\ee
In order to prove that the two scalars $A(x)$ and $C(x)$ are the same it is sufficient to prove this in a particular gauge, 
order by order in perturbation theory.
We find convenient to choose the UCG, as before. We shall see that indeed $A^{(1)}(x)=C^{(1)}(x)$ and $A^{(2)}(x)=C^{(2)}(x)$.
The first order quantities are easily analysed since
\be 
A^{(1)} = \dot{\phi} \int^t dt' \alpha = \varphi = C^{(1)}\,,
\ee
so that the equivalence is proved at first order.
At second order, plugging the choice  $A^{(0)}(t)=\phi(t)$ into Eq.(\ref{A_SG}), considering the LW limit approximation and using the results of the previous section, we have 
\be 
A^{(2)}=\left(\frac{1}{2}\frac{\ddot{\phi}}{\dot{\phi}^2}+\frac{1}{4 M_{Pl}^2} \frac{\dot{\phi}}{H}\right) \varphi^2\,.
\ee
On the other hand, using Eq.(\ref{Lastphi2}), and performing the integral in the LW limit, we find
\be 
C^{(2)}=\varphi^{(2)}=\left(\frac{1}{2}\frac{\ddot{\phi}}{\dot{\phi}^2}+\frac{1}{4 M_{Pl}^2} \frac{\dot{\phi}}{H}\right) \varphi^2\,,
\ee 
which means that the equivalence is also true at second order.
Then, as anticipated, the UFG observers are physically equivalent to the free-falling ones and they experience the same backreaction.

\section{Conclusions}
In this paper, 
we have extended to an arbitrary potential, which specifies a single field inflationary model, and to all orders in the slow-roll parameters
a previous investigation~\cite{FMVVback} devoted to the computation of the quantum corrections due to scalar cosmological fluctuations
on the effective expansion rate and equation of state as seen by a free falling class of observers.
This problem is also known as the quantum backreaction problem which has been widely investigated in the literature.

We have worked in a cosmological perturbation theory setting, up to second order, and in a consistent genuinely gauge invariant approach \cite{GMV1,GMV2}. 
This means that the observable, which depends on a class of privileged  observers, once defined, can be computed in any coordinate system. 
The result that we obtain tells that a free-falling observer does not see any change in the effective expansion rate and equation of state, 
if we take into account all the LW scalar fluctuations of the inflaton and of the metric in the semiclassical approach. 
Namely, for any given potential of the inflationary model, the homogeneous background value of the Hubble factor coincides with the effective Hubble factor 
(seen by the free falling observers) when the scalar LW quantum fluctuations (up to second order) are turned on.
Moreover, this class of observers, even in the presence of the quantum fluctuations, is perfectly equivalent to the one comoving with the inflaton field,
that is the class of observers who see the inflaton as homogeneous. As a consequence, these last observers share the same physical properties of 
the free falling observers and they do not see any backreaction effects.

It would be interesting to apply similar methods to the analysis of a pure gravitational $R^2$ model \cite{Starobinsky:1980te}
(where also the tensorial fluctuations can play a role in the backreaction~\cite{FMVVgravitons})
and to Higgs inflation models like \cite{Bezrukov:2007ep,Barvinsky:2009ii}, which look particularly appealing after the release of the recent Planck results (see, in particular, \cite{Ade:2013uln}).\\


\vspace{0.5cm}
\noindent

{\bf Acknowledgements}

We would like to thank Giovanni Venturi for useful discussions.
GM is supported by the Marie Curie IEF, Project NeBRiC - ``Non-linear effects and backreaction in classical and quantum cosmology".


\end{document}